\documentclass[prl,aps,twocolumn]{revtex4}
 \pdfoutput=1
\usepackage{graphics}
\usepackage[]{color}

\begin{document}
\definecolor{magenta}{rgb}{1,0,1}
\definecolor{555}{rgb}{0.5,0.5,0.5}
\definecolor{999}{rgb}{0.9,0.9,0.9}
\definecolor{0095}{rgb}{0.,0.,0.95}

\title{
Can humans see beyond intensity images?}

\author{Geraldo A. Barbosa$^*$}
\affiliation{
\mbox{Northwestern University - Department of Electrical Engineering and Computer Science},
2145 N. Sheridan Road, Evanston, IL 60208-3118}
\affiliation{\mbox{QuantaSec -- Consulting and Projects in Physical Cryptography Ltd.}\\
Av. Portugal 1558, Belo Horizonte, MG 31550-000, Brazil.  \\ $^*${\em \small GeraldoABarbosa@gmail.com}}

\date{Submitted: Feb 2012}

\newcommand{\be}{\begin{equation}}
\newcommand{\ee}{\end{equation}}
\newcommand{\bea}{\begin{eqnarray}}
\newcommand{\eea}{\end{eqnarray}}

\begin{abstract}
The human's visual system detect intensity images. Quite interesting, detector systems have shown the existence of different kind of images. Among them, images obtained by two detectors (detector array or spatially scanning detector) capturing signals within short window times may reveal a "hidden" image not contained in either isolated detector: Information on this image depend on the two detectors simultaneously. In general, they are called "high-order" images because they may depend on more than two electric fields. Intensity images depend on the square of magnitude of the light's electric field.  Can the human visual sensory system perceive high-order images as well? This paper proposes a way to test this idea. A positive answer could give new insights on the "visual-conscience" machinery, opening a new sensory channel for humans. Applications could be devised, e.g., head position sensing, privacy in communications at visual ranges and many others.
\end{abstract}

\maketitle


\section{Introduction}
The quest discussed in this work touches at the interface of our understanding of the visual processing system and the notion of conscience.
 Understanding how human vision works was greatly advanced by the discovery of light-activated channels, e.g, the discovery of conformational changes of ``light pigments'' (rhodopsin and photopsins) within rods  and cones --the two types of photoreceptors in the human retina.  These changes modify the ionic permeability   of the cell membrane, transducing light into electrical signals that travel  to the retinal ganglion cells  and to the optic nerve. At the optic chiasm, these nerve fibers are divided to each side of the brain and
 reach the primary visual cortex. Visual ``conscience'' is formed from correlations among a multitude of recorded signals \cite{Kenyon}. The understanding of these correlations is an open question.  From another side, advances have been made in the study of quantum aspects of light  \cite{MandelWolf,Glauber} that led to the discovery of ``high-order'' images \cite{HighOrder,Ribeiro}, revealed by detectors recording signals within short window times (``time-coincidence'' detections). These images will be discussed ahead but just to mention a simple case, two detectors are probing (e.g., by scanning)  light intensity from a light source and each one, independently, sees just an intensity blur. Preparing the detectors so that their signals are taken within short time windows (a time-correlated detection) and only accepting them when both detectors are activated simultaneously, the blur signal disappear and a completely different image may appear, e.g. a letter ``A''. This image ``A'' is {\em not} present in either detector but depends on {\em both} detectors. It is natural to ask if  similar ``higher order'' correlations may exist in the brain as a mechanism to combine multiple received signals and generate a ``high-order'' image.
  The answer to ``Can humans see beyond intensity images?'' involves problems at the frontiers of biological and exact sciences.
    In this paper an experimental strategy is proposed that would enable us to answer this fundamental question about human vision.

  Several difficulties stand in the path to obtain this answer. Among them, the efficiency of converting light to electric signal is not high ($\sim 0.5$) and the information transfer is further degraded by neural noise in the spike-trains \cite{Troy}. This noise is the ultimate limiting factor defining the threshold for light sensitivity in humans ($\sim 7$ or $8$ photons) \cite{Hecht,Sakitt}.
 Spike trains ($ \sim 2 \times 10^1$ to $4 \times 10^2$ ms in duration) produce the spatiotemporal retinal images. Experimental conditions have to accommodate such time durations. Nonlinear processes also exist such as nonlinear interactions at the cones regulating the variable sensitivity of the visual system \cite{Troy,Levine}. Recording what is a spatiotemporal image involves a sequence of nonlinear operations enriched with feedback control loops \cite{Hateren}. The number of neurons involved is very high, stressing the computational resources to simulate signal correlations.
 Nevertheless, no in-principle arguments seems to exist against obtaining a clear answer for this fundamental question. Besides philosophical aspects, there are also practical aspects involved.
 Use of infrared or ultra-violet vision equipment already revolutionized our understanding of the world. They gave, among others, glimpses of how other creatures, e.g., insects, may view the world. Images beyond intensity images enters in a different and broad category, yet to be explored.



\section{Methods}

\subsection{Light Sources}
The experiment to be described involves a special single  light source, of low intensity, that emits in a somewhat broad spatial range so that light could reach both eyes of the subject. Light from this light source carries the desired information but not each separate beam reaching either eye. The desired information is inherent to both beams or, in other words, it exists in the correlation of the two beams.

\begin{figure}
\centerline{\scalebox{0.42}{\includegraphics{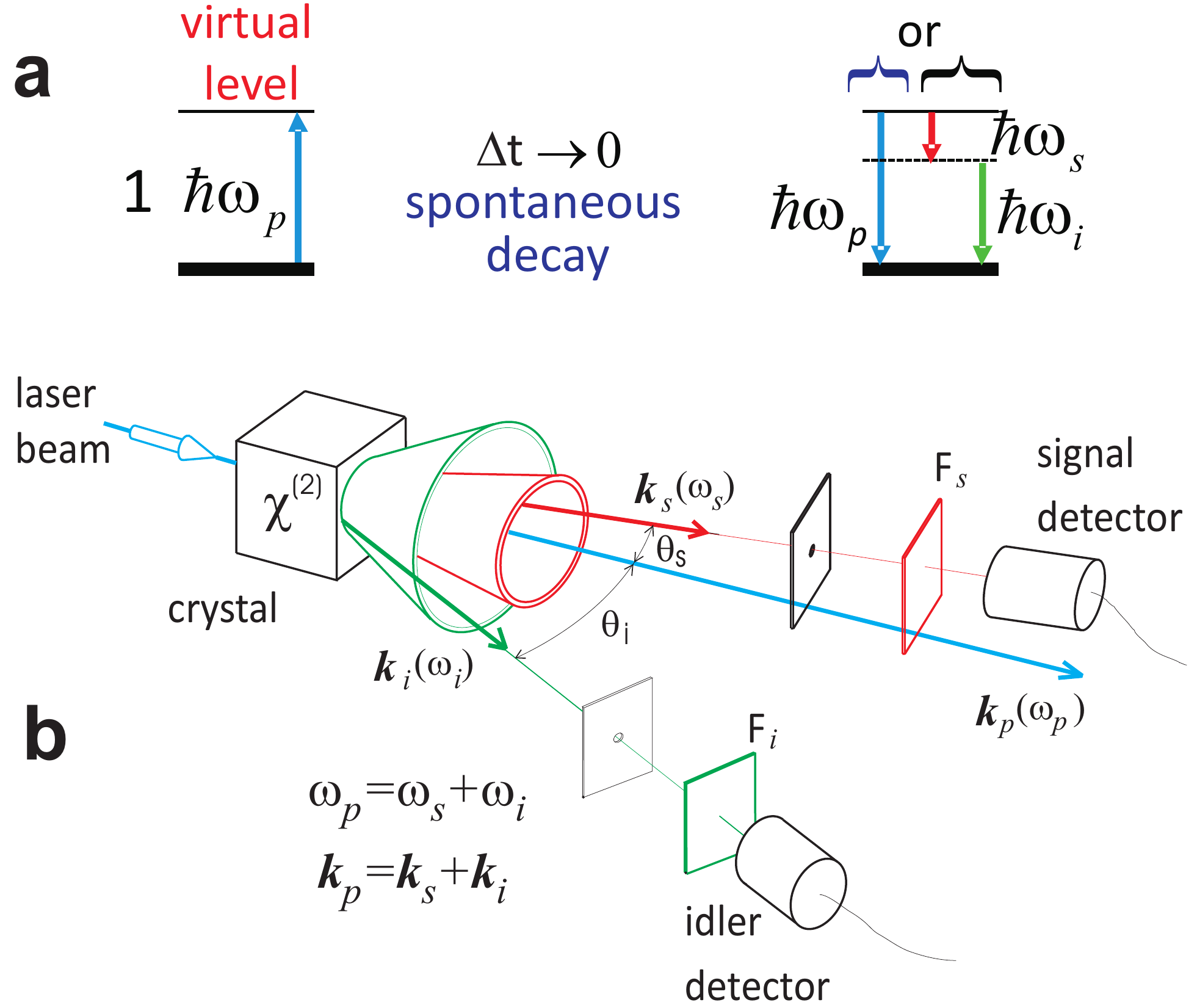}}}
\caption{
\label{Barbosa_fig1}
{\bf a}) Spontaneous parametric down conversion: A photon from a laser beam excites a nonlinear medium (usually a crystal) to a virtual level that decays spontaneously into two possible paths. Either an identical laser photon is created or a pair of photons appear.  Their energy adds up to the energy of the laser photon ($\hbar \omega_p=\hbar \omega_s+\hbar \omega_i\rightarrow$ energy conservation). Photons from all colors (wavelengths) can be obtained.
{\bf b}) The photon pairs follow directions given by linear momentum conservation ($\hbar {\bf k}_p=\hbar {\bf k}_s+\hbar {\bf k}_i)$. $F$ designate an optical interference  filter.
}
\end{figure}

This light source is mainly created by nonlinear crystals  excited by laser light.  A nonlinear crystal refers to a crystal made from a nonlinear medium or, in other words, a medium in which the electric field of light induces nonlinearly a polarization of that medium. Usually this nonlinearity occurs at very intense values of the electric field so that the atoms in the medium are displaced at positions beyond their normal ``harmonic'' oscillatory positions.
 The process of converting laser light to the special light states needed is known as {\em parametric down conversion}  (PDC) \cite{BandW}.
 Emissions from this excited crystal, at low intensity, usually occur in photon pairs that can be detected at quantum level, photon by photon. Detectors with single-photon sensitivity are used. At these intensity levels, signals are below the human eye sensitivity. When no other source of excitation for the crystal is present, besides the pump beam,  the  phenomenon is known as {\em spontaneous} parametric down conversion (SPDC). Every detection of a photon from a pair can be precisely space-time correlated to the emission of its conjugate photon --See Fig.~\ref{Barbosa_fig1}.
 Photons in these pairs are traditionally called signal and idler photons. These emissions occurs at random times.
The energy and momentum conservation laws for the light interaction with the nonlinear medium lead to the ``phase matching conditions'' \cite{PhaseMatch}: They define emission angles and wavelengths for the signal and idler photons.

The laser interaction with the nonlinear crystal and the resulting light states are known through the Hamiltonian of the SPDC phenomenon. The Hamiltonian describes, in a quantum way, the light-matter interaction and the resulting state produced by this interaction. This state, also known as wave state or wave function, gives the maximum information available about a physical system (See, for example, \cite{BandW,Barbosa-WFunction}).

\subsection{High-Order images}
The literature on ``high-order'' images (loosely also called  ``quantum images'')  is quite varied; e.g., see \cite{QuantumImaging}. A few simple examples of ``high-order'' images will be shown to introduce the reader to this concept and to give a glimpse of some basic features --the most basic one being the dependence from information coming from, at least, two detectors.

\begin{figure*}
\centerline{\scalebox{0.95}{\includegraphics{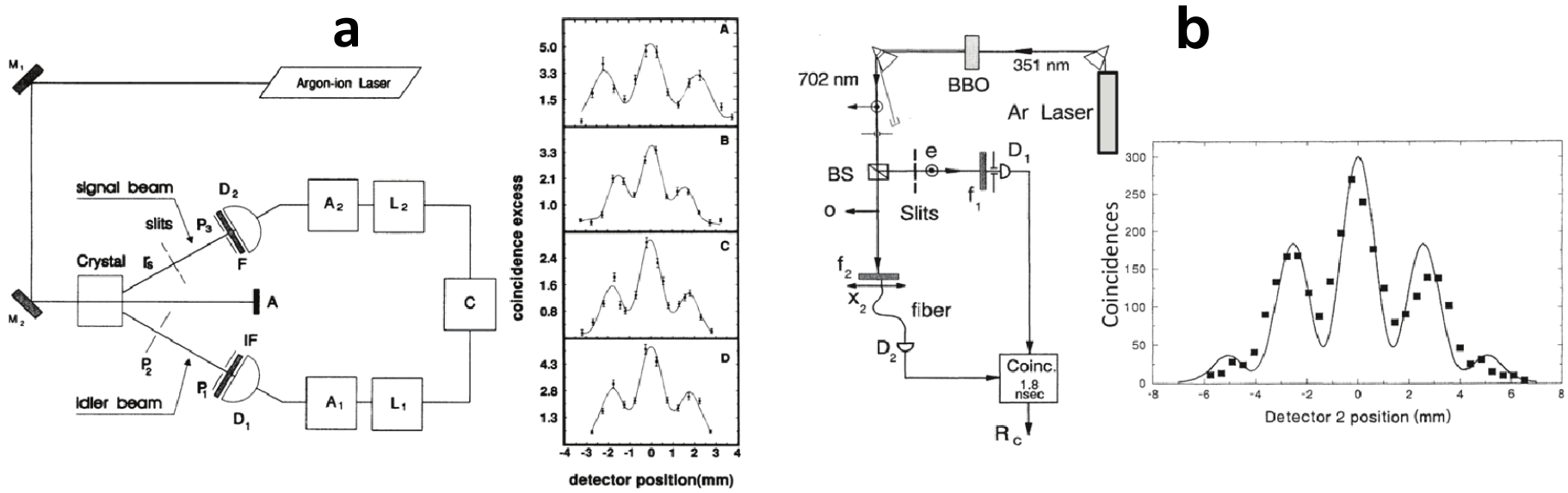}}}
\caption{{\bf a}) Left: A Young double-slit mask is inserted in the signal beam path of the SPDC. The position of the mask can be set so that the separation of the slits is much {\em larger} than the coherence area of the signal beam. Classically, this condition defines that {\em no} interference fringe could be seen.
 M are mirrors, D are detectors, IF are interference filters, A, L and C are parts of the electronics coincidence system. Right: Detected coincidence fringes are seen even when the mask is placed directly onto the crystal (light source).
  {\bf b}) Left: Setup for detection of coincidence fringes. {\em Collinear} SPDC: the two down converted beams are collinear with orthogonal polarizations. Beams are separated by the beam splitter BS. The two-slit pattern is {\em not} on the scanned beam but along the path going to the fixed detector. Right: ``Ghost'' interference and diffraction pattern.
\label{Barbosa_fig2}
}
\end{figure*}

The very first experimental detection of a higher order spatial image was an interference fringe from a double-slit mask. The central point was that the experimental arrangement was such that,
 classically, {\em no} interference fringe could be seen \cite{Ribeiro}: the slit separation was much larger than the light's coherence area at that position (Fringes are only seen when the electric field at both slits vary coherently).  The experimental setup used and the obtained results are shown in Fig.~\ref{Barbosa_fig2}-A (right).
The obtained pattern in Fig.~\ref{Barbosa_fig2}-{\bf a} was obtained accumulating data at short time windows when the two detectors clicked ``simultaneously'' --one says that the detectors were in a ``coincidence'' operation.
The detected coincidence pattern establish that the appearance of interference fringes depends simultaneously on  information carried by {\em both} beams but do not exists in a single beam.  As the distance $r_s$ from the crystal to the double-slit mask varies, the fringe visibility has a very small variation. In case (d), the double slit is touching the crystal: The signal beam coherence area is too small to allow any fringe to be seen (See \cite{RibeiroMonkenBarbosa}).
The setup used still allows classical fringes to be seen {\em if}  $r_s$ is set suitably large to make the coherence area larger than the slits separation.

 The second example, Fig.~\ref{Barbosa_fig2}-{\bf b}, taken from \cite{StrekalovShih}, demonstrates the same phenomenon but in a condition where interference fringes could not have been seen anyways using detections of a single beam. The reason is simple: the double-slit mask was {\em not} on the signal beam path being scanned, but on path of the fixed detector along the idler beam.
This experiment is, despite geometric differences, intrinsically similar to the first one described. They emphasize that the fringe information depends on {\em both} beams, differently from a classical interference pattern \cite{Barbosa96}.

Many other examples can be found in the literature, but these two simple historical experiments are sufficient to stress the existence of high-order images. These  examples have shown  correlations \cite{correlation} between {\em two} simultaneous ``intensities''.  The normal intensity image depends on the expected value of  $\langle E({\bf r},t)^- E({\bf r},t)^+ \rangle$ \cite{Glauber}(classically $\rightarrow \langle |E({\bf r},t)|^2 \rangle$), where $E$ is the electric field operator, at a single space point.  High-order images involve more electric fields at general $({\bf r},t)$ positions, e.g., $\langle E({\bf r}_1,t_1)^- E({\bf r}_2,t_2)^-\cdots  E({\bf r}_2,t_2)^+ E({\bf r}_1,t_1)^+ \rangle$ \cite{Glauber}.  Other possibilities exist for generation of high-order images. Some of them rely on purely quantum aspects of light --such as the ones that need twin photon states-- while others can be created even with standard sources of light (sometimes called sources of ``classical'' light, e.g. sunlinght).
\subsection{Pattern transfer from the pump beam}
The examples given described high-order images produced by masks placed on the path of a down converted beam. Other arrangements produce high-order images as well. To exemplify one possibility, related to the experiment to be proposed, one could imprint a given pattern directly on the intensity of the pump beam. The experiments already described  have all utilized a pump beam where the beam mode intensity had a spatial Gaussian profile. One could even consider the beam  profile itself (say, a Gaussian shape) as the information to be transmitted --or any other geometric feature that could be carried by the laser beam.

Consider a laser beam with a Gaussian intensity profile. The down-converted light presents a continuum of wavelength possibilities, in a rainbow of colors, regardless the laser profile being used. Therefore, the measured intensity produces a smooth intensity on all places where the phase matching condition allows light to be present. Intensity measurements  at the phase matching positions do {\em not} reveal the Gaussian pump profile. However, with coincidence detection keeping one of the detectors at a fixed point while scanning the other detector, that smooth background disappears and the spatial Gaussian-like profile of the pump beam is revealed --as a high-order image.

Of course, different information can be imprinted on a laser beam and can, in principle, be transmitted through detection of signal and idler photons.
As an example, a stop-mask where a letter has been cut to allow full transmission, will appear in the coincidence detection scheme, but {\em not} in direct detection of the intensity of either down-converted beams. See \cite{Monken,Koprulu}.
\begin{figure*}
\centerline{\scalebox{0.7}{\includegraphics{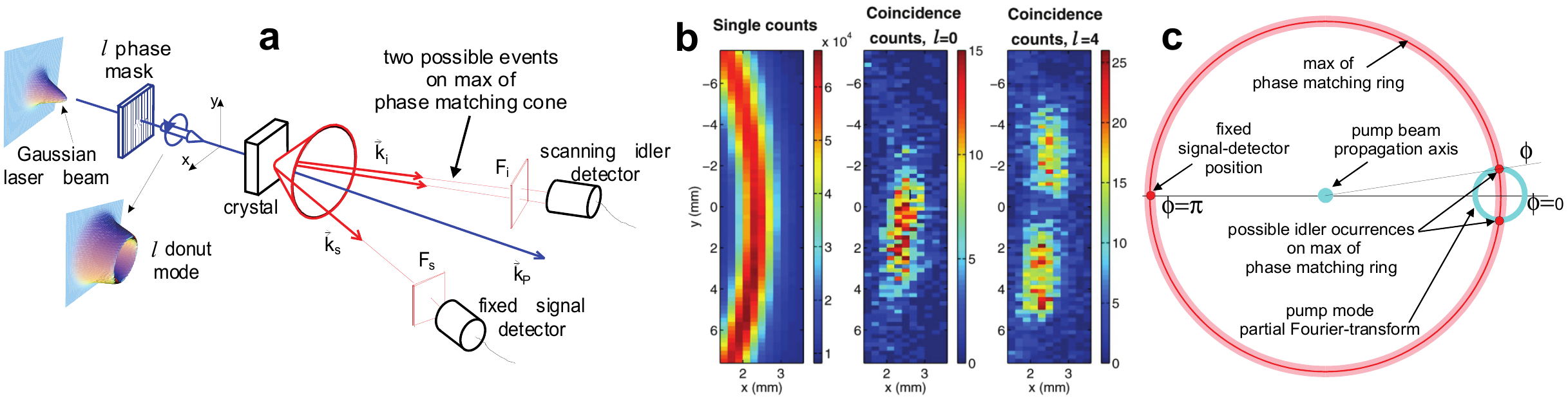}}}
\caption{{\bf a}) Experimental setup. A
laser beam with wave vector ${\bf k}_P$ ($\lambda=351$ nm) is
prepared in the $l=4$ state of orbital angular momentum
and focused on a $\chi^{(2)}$ crystal (type-I cut
phase matching). F designate interference filters centered at 702nm.
{\bf b})
Left: Single counts for idler: This is the kind of intensity image that could be seen by a single eye; Center: coincidences counts for
pumping with a Gaussian beam ($l=0$): a high order image; Right:  coincidences for a pump beam with
$l=4$: a high order image. {\bf c})  A detection-plane representation of the
pump's donut-mode transform (See \cite{Barbosa-WFunction}) intersecting the
down-conversion ring. It presents two crossings at the max of the phase matching condition, as measured in {\bf b}), with $l=4$.
\label{Barbosa_fig3}
}
\end{figure*}
As another example, Fig.~\ref{Barbosa_fig3}-{\bf a} shows the setup and results of an experiment \cite{Altman} where the laser beam mode has a  ``donut-shape'' profile (consider it as the information to be extracted) and a special phase signature producing a beam carrying {\em orbital}  angular momentum $l$: the Poynting vector carrying the mode energy  circulates around the beam's axial direction \cite{Allen,Barnett}.  Orbital angular momentum modes provide an easy way to imprint ``donut'' profiles onto a light beam. To do so, special optical masks, holographic filters, adaptive pixel reflectors,... are used. Moreover, whenever a pump laser with an orbital angular momentum mode produce a pair down-converted photons, the angular momentum may be transferred to the photon pair (angular momentum conservation) \cite{HugoBarbosa,Zeilinger}.  The results are shown in Fig.~\ref{Barbosa_fig3}-{\bf b}.

The results shown were acquired in a geometry that used a lens properly focused and interference filters to produce a narrow down conversion bandwidth. A full donut would appear with a broad down conversion bandwidth.
See \cite{woerdman} and \cite{FengChenBarbosaKumar}.

The main question could be refined to ``Can the Eye-Brain system perform time and space correlations necessary to reveal high-order images?''.
\begin{figure}
\centerline{\scalebox{0.5}{\includegraphics{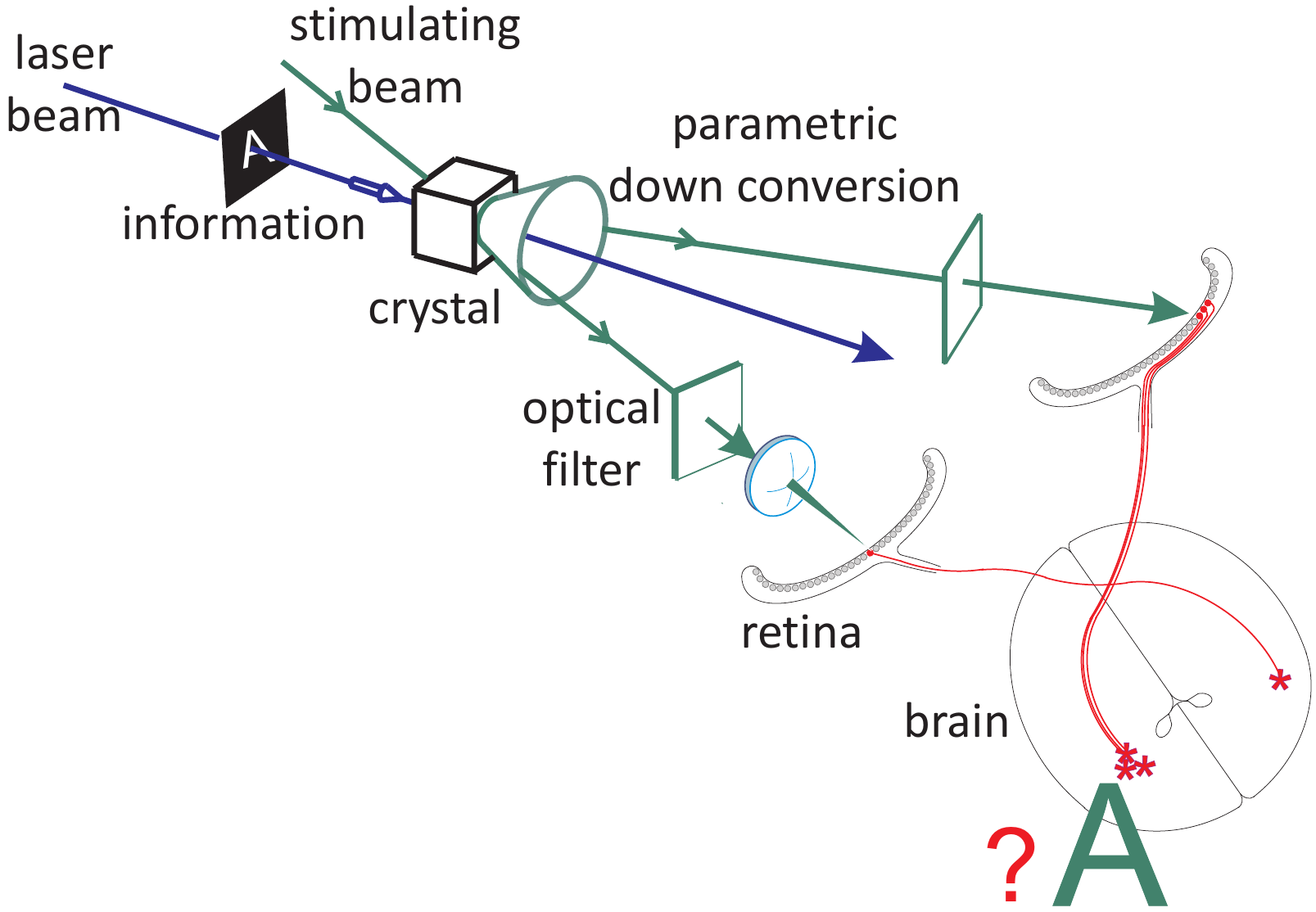}}}
\caption{Sketch of a setup to recover the information imprinted on a laser beam (a letter A, in this example) by visualization of a high-order image.
\label{Barbosa_fig4}
}
\end{figure}

\subsection{A proposed experiment}
Fig.~\ref{Barbosa_fig4} sketches a proposed setup. A specific information was imprinted on a laser beam. Basically, what is being thought is directing signal and idler beams to each eye and ask if correlations in the brain will allow visualization of a high-order image (intensity-intensity correlation).
Any detail related to the optical systems needed to direct these beams to the two eyes are neglected at this moment.

\subsection{Some expected difficulties}
Several impairments easily appear against a positive answer to the question whether the Eye-Brain system could reveal high-order images.

The first one is that the SPDC efficiency is very low. Signal and idler photons mainly appear in single pairs: SPDC has negligible probability to produce multiple photon pairs within a coherence time. However, human vision sensitivity --within the visible range-- only works starting with ~7 photons and not less.  Besides this, the time response $\tau_r$ to produce retinal images demand at least $\tau_r \sim 20$ms. With common laser pump intensities in continuous operation, in this kind of time intervals a high number of photon pairs can be produced --each within short window times $\ll \tau_r$. Despite that $n_s$ and $n_i$ photons are highly correlated in each emission, photons in distinct emission times have no correlation between them. Thus all ``good'' correlations existing in every emission could be washed out by the large number of uncorrelated pair events occurring in $\tau_r$ and, together with them, the high-order images. Therefore, one of the main problems is if there is a way to produce in  a single shot a correlated number of photons sufficient to produce a retinal signal at or above the minimum threshold of 7 photons within a sampling time of order $\tau_r$. Repetitions are allowed at intervals above $\tau_r$.

Another possible deleterious problem is that signals generated by a group of cones may be directed to a horizontal cell in a data binning process \cite{SmithEtAl} and some information may be lost in this process. Sensitivity regulation of raw signals may be occurring in feedback processes whose effects are not clear concerning conservation of correlation properties among different inputs.

These questions do not have a clear answer at the moment and become aspects to be analyzed both experimentally as well as theoretically.

\subsection{Stimulated emission}
Stimulated emission is a process that enhances parametric emission, providing more photons than SPDC.

SPDC implies that the initial quantum state of light in the down-converted states, $|\Psi (0)\rangle_{\tiny signal}\otimes |\Psi \rangle_{\tiny idler}$, is just the vacuum (state of light with no photons present): $|0 \rangle_{\tiny signal} |0 \rangle_{\tiny idler}$.
The function describing the SPDC process will be given by a time evolution from initial state of the system. In SPDC, this initial state will be the vacuum state.

In {\em stimulated} light emission, the initial down-converted state is {\em not} the vacuum but another light state. It can be produced by an auxiliary or stimulating  Gaussian laser beam directed along the path, say, of the signal beam (See Fig~\ref{Barbosa_fig4}). As soon as a pump photon excites the nonlinear medium, the electric field associated to the stimulating beam induces a decay of the excited medium. Both signal and idler beams are enhanced: The light levels may fall onto the visible range.
 For a nonzero vacuum state as the initial state for the signal beam, the resulting state at an arbitrary instant $t$ is given by the evolution
\bea
\label{Psi}
|\Psi(t) \rangle=e^{  -i\int^t_{t-t_{int}}H dt   }|\Psi(0) \rangle_{\tiny signal} |0 \rangle_{\tiny idler}\:.
\eea
The Hamiltonian operator can be written in terms of the photon annihilation and creation operators for signal and idler photons ($a,a^+$ and $b,b^+$) as $H=(\kappa\: a^+ b^+ +\kappa^*  b a)$, where $\kappa$ includes the crystal efficiency and the pump laser amplitude $\alpha_p$,  taken as an intense classical field  (see Appendix).

 For this work, an algebraic expansion of (\ref{Psi}), using the signal initial state as a coherent state \cite{Glauber}, $|\Psi(0) \rangle_{\tiny signal}=|\alpha \rangle$, was done for arbitrary number of signal photons $n_s$ and up to ten idler photons $n_i$. The magnitude square of the coefficients in this expansion determines the probability $P(n_s,n_i)$ to find $n_s$ {\em and} $n_i$ photons, within a short sampling time
 $\Delta t\ll\tau$, in the down-converted light (See Appendix). The laser coherence time $\tau$ can be short or long, depending on the laser cavity used (e.g., from  $\tau \sim 10^{-10}$ to $10^{-3}s$).

\begin{figure*}
\centerline{\scalebox{0.8}{\includegraphics{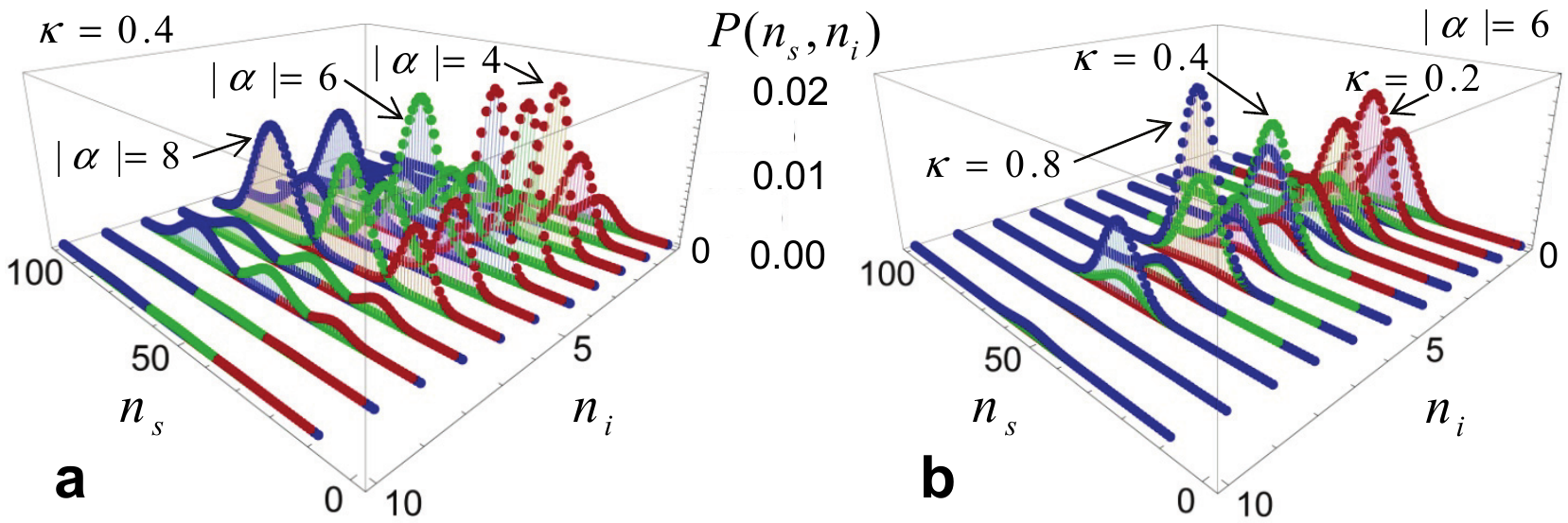}}}
\caption{{\bf a}) $P(n_s,n_i)$ for simultaneous occurrence of $n_s$ and $n_i$ photons. For a coherent field  $\langle n \rangle=|\alpha|^2$ \cite{Glauber}. Parameters $|\alpha|=8,6,2$ and $(|\kappa|=0.4;\phi=0)$.
{\bf b}) $P(n_s,n_i)$ with parameters $(|\kappa|=0.8,0.6,0.2;\phi=0)$; the laser amplitude is $|\alpha|=6$.
\label{Barbosa_fig5}
}
\end{figure*}
 Fig.~\ref{Barbosa_fig5}-{\bf a} shows that as $|\alpha|$ varies, peaks mainly displace along the $n_s$ axis. In contrast, Fig.~\ref{Barbosa_fig5}-{\bf b} shows that as $|\kappa|$ varies, peaks displace along the $n_i$ axis. This indicates $|\alpha|$ and $\kappa$ as two parameters for experimental control. While $|\alpha|$ is just controlled by the auxiliary laser intensity, $\kappa$ depends mainly on the crystal used (susceptibility and crystal length) as well as on the pump laser focus being used (See \cite{Barbosa-WFunction}).

 $P(n_s,n_i)$ applies for photon number measurements within one laser coherence time. Each spot represents a possible decay from the crystal excited by a single laser photon.
 One may assume that the pump laser operates in a flash mode with duration of $\Delta t \ll  10^1$ms (Human response time), but in a slower rate.

Distinctly from SPDC,
stimulated down-conversion does not present signal and idler photons with balanced numbers (or, $n_s=n_i=1$). See  Fig.~\ref{Barbosa_fig5}.
 In this case,  more photons go towards one eye than the other one in Fig.~\ref{Barbosa_fig4}. It is not clear if this is deleterious for high-order image detection,  as the available experiments have been done with single photons. However, as the wave function (\ref{Psi}) contains all information available, a stronger signal beam will also carry the necessary information.  The imprinted information could be the total orbital angular momentum $l$ (it gives a particular donut shape). In this case, for a signal beam with $l_s=0$, such as a Gaussian auxiliary laser in the signal path, the idler photons should carry the total momentum $l_i=l$ (angular momentum conservation) \cite{HugoBarbosa}.

\section{Discussions}

 The stimulated emission technique creates light intensity conditions to allow human beings perceive high-order images. However, whether the brain is able to acquire spatial information even with photon number unbalance is to be tested experimentally. Another unbalance effect is that when  $n_s$ photons excites one retina, the number of photons exciting the other retina may be divided into two classes: one below the retina sensitivity threshold ($\leq 6$) and the other above it. This threshold effect produces a further signal unbalance in the neural paths
 (high-order image on a brighter background?).

Computer simulations may provide insights into these complex problems.

 Recent advances described the  transfer functions for the signal channels from the retina to the axions (e.g., \cite{Hateren,Troy}).
Nonlinearity associated with regulatory feedback systems simulated real signals. Transfer functions were used to simulate formation of intensity images in the brain \cite{Kenyon}.
 Simulation of high-order images may demand resources up to $\sim N(N-1)/2\sim N^2/2 \times$  the resources necessary to simulate a $N$-``pixels'' intensity image \cite{Kenyon}. This is a challenging but very rewarding task.

Other stimulating setups can be devised and studied, either theoretically or experimentally. For example, two stimulating beams can be used, one for each ``signal'' and ``idler'' paths.  While one may say that both down-converted beams will get stronger, it is known that photons in a coherent state do not have number correlation between them: $\langle n_1 n_2 \rangle=\langle n_1 \rangle \langle  n_2 \rangle$. Even if the desired imprinted information from the laser beam is carried by the signal and idler photons, this lack of correlation will probably lead to a strong uncorrelated background in both eyes.

\subsection{Illuminating optical system, laser and crystal specifics}

 One should also specify the necessary basic equipment for this kind of experiment. As a pump laser, some commercial diode pump lasers operate at $\lambda_{pump}=266$nm.
    For degenerate PDC, $\lambda_{pump}$ produce signal and idler photons at $\lambda_s=\lambda_i=524$nm --at the crossover of the photopic and scotopic eye sensitivities. Increasing laser coherence length,  producing pulsed operation or finding an auxiliary laser to stimulate $n_s$ decays are just technicalities.
    Nonlinear crystals  to work in this  UV range are, e.g., Beta Barium Borate (BBO), Lithium Triborate (LBO) and CBBF ($CsBe_2BO_3F_2$); a variety of periodically pulled nonlinear crystals (PPLN) crystals can be also explored.

Special goggles to direct photons to precise regions of the human retina --e.g., densely packed cone region-- have to be designed.
Precise eye-attractor points have to be set directing vision towards a reference point while the light under analysis strike the retina at a fixed angle away from that reference.

 The light levels from the signal and idler photons involved offer absolutely no risk to human subject -- they are much lower than the residual light in a dark bedroom at night.

 One could observe that $\kappa$ was taken as a complex parameter $\kappa=|\kappa|\exp(i \phi)$, with $|\kappa|$ and $\phi$ considered real. Actually, a more careful look at  $\kappa$ shows a geometric structure in the $(k_{sx},k_{sy};k_{ix},k_{iy})$ coordinate system that represents the high-order image to be seen in a coincidence experiment \cite{Barbosa-WFunction}. This structure is implicit in $\widetilde{\psi }_{lp}\left( \Delta {\bf k}\right)$ (See Appendix).

Detection of high-images requires that obstacles that could clip any of the light beams from the nonlinear crystal to the retina should be avoided \cite{Indistinguishability,cryptoCut}.

These experiments could eventually be complemented with  functional magnetic resonance imaging (fMRI) as long as signals are strong enough to produce a hemodynamic response within a few seconds ($\sim 7$s).

\section{Conclusions}

The experiments sketched in this work intend to reveal if humans can use their visual sensory system in a way never appreciated before. Other suggestions may stimulate interesting debates on this subject.
 In principle, the outcome of a given experiment will give a simple ``yes'' or ``no'' answer about the success of seeing a high-order image (such as a letter or a generic symbol --otherwise invisible as an intensity image) under the conditions used. A positive answer would indicate an extension of the information processing capacity of humans --a worthwhile endeavor.

\section{Acknowledgments}
The author thanks
 Todd Parrish, David Ferster,  Sheng Feng and Prem Kumar (Northwestern University) for helpful discussions on potential difficulties that may be expected in these experiments, and Daniela Drummond-Barbosa (Johns Hopkins University) for a critical reading of the manuscript.

\section{Appendix}

The wave function derived from the interaction Hamiltonian for PDC, for the stimulated case, can be written in a simplified compact form as
\bea
\label{Psi}
|\Psi \rangle=e^{ - i  \int_{t-t_{int}}^{t} H dt      }|\Psi \rangle_{\tiny signal} |0 \rangle_{\tiny idler}\:,
\eea
where \cite{MandelWolf,Barbosa-WFunction}
\begin{eqnarray}
\label{H}
{H} &=&\!\sum_{{\sigma} ,{\sigma} ^{\prime }}\!\int \!
d^3k\:d^3k^{\prime} \:
l_{  \omega_{\bf k}     }^{(*)}
l_{\omega_{\bf k}^{\prime}}^{(*)}
{a}^{\dagger}\!\left( {\bf k},{\sigma} \right)
{b}^{\dagger}\!\left( {\bf k}^{\prime},{\sigma}^{\prime}\right)\nonumber\\
&&\times e^{    i   \left(    \omega_{\bf k}+\omega_{   {\bf k}^{\prime}    }      \right) t     }
 \chi _{qjk}
\left( {\bf e}_{   {\bf k},\sigma                   }\right)_j^{*}
\left( {\bf e}_{   {\bf k}^{\prime},\sigma^{\prime} }\right)_k^{*}\nonumber\\
&&\times \int_{V_I}\!\!d^3r\left( {\bf E}_{P}\left( {\bf r},t\right)\right)_q
e^{   -i\left(   {\bf k}+{\bf k}^{\prime}    \right) \cdot {\bf r}    }\!\!+\!h.c.\:\:.
\end{eqnarray}
(Here, signal and idler indexes ($_s,_i$) are replaced by non-primed and primed variables).
  ${\bf E}_P\left( {\bf r},t\right) $ is the electrical field associated with the pump beam (it may be assumed classical for strong fiels),  $ {\bf k}$ and $ {\bf k}^{\prime}$ indicate signal and idler wave vectors and indexes $(\sigma,\:\sigma^{\prime})$ represent possible polarization states. $\chi$ is the electric nonlinear susceptibility tensor.
  Repeated indexes $q$, $j$ and $k$ imply summations.

  Restricting the analysis for wave vectors very close to the maximum of the phase matching \cite{PhaseMatch} and ignoring the explicit polarization indexes $\sigma$s, the Hamiltonian could be written
\bea
{H}_{pm} &\simeq&\!\!\sin \theta'_{pm} k_{pm}^{\prime 2}\sin \theta_{pm} k_{pm}^2
l_{  \omega_{{\bf k}_{pm}}     }^{(*)}
l_{\omega_{{\bf k}_{pm}}^{\prime}}^{(*)}
{a}^{\dagger}\!\left( {\bf k}_{pm} \right)
{b}^{\dagger}\!\left( {\bf k}_{pm}^{\prime}\right)\nonumber\\
&&\times e^{    i   \left(    \omega_{{\bf k}_{pm}}+\omega_{   {{{\bf k}}^{\prime}_{pm}}    }      \right) t     }
 \chi _{qjk}
\left( {\bf e}_{   {{\bf k}_{pm}}            }\right)_j^{*}
\left( {\bf e}_{   {{\bf k}^{\prime}_{pm}} }\right)_k^{*}(\Delta k)^6\nonumber\\
&&\times \int_{V_I}\!\!d^3r\left( {\bf E}_{P}\left( {\bf r},t\right)\right)_q
e^{   -i\left(   {\bf k}_{pm}+{\bf k}_{pm}^{\prime}    \right) \cdot {\bf r}    }\!\!+\!h.c.\:\:,
\eea
where $a$ and $b$ are annihilation operators for signal and idler photons. The range of wave vectors that contain the desired structure information, close to the maximum for phase matching, can be calculated as shown in Ref.~\cite{Indistinguishability}. For the example given in Ref.~\cite{Indistinguishability} the wave vectors correspond to polar and azimuthal angular variations of order $\Delta \theta\sim \Delta \varphi\sim0.02$rd around the optimum position.

Neglecting the $_{pm}$ indexes, the term
$- i  \int_{t-t_{int}}^{t} H dt$ is written
\bea
\label{pm}
&&- i  \int_{t-t_{int}}^{t} H dt \rightarrow -i\sin \theta'\: k^{\prime 2}\:\sin \theta \:k^2\:
l_{  \omega_{{\bf k}}     }^{(*)}\:
l_{\omega_{{\bf k}}^{\prime}}^{(*)} \nonumber \\&&\times  \chi _{qjk}
\left( {\bf e}_{   {{\bf k}}            }\right)_j^{*}
\left( {\bf e}_{   {{\bf k}^{\prime}} }\right)_k^{*}
T(\Delta \omega) \:
\widetilde{\psi }_{lp,q}\left( \Delta {\bf k}\right)\nonumber \\&&
{a}^{\dagger}\!\left( {\bf k} \right)
{b}^{\dagger}\!\left( {\bf k}^{\prime}\right)(\Delta k)^6
 \!\!+\!h.c.\:,
\eea
where\\
$T(\Delta \omega)$$=$$\exp\left[i\Delta \omega
\:(t-t_{int}/2)\right]\:$$ \sin\left(\Delta \omega\:
t_{int}/2\right)/\left(\Delta \omega /2\right) $ is the time window
function defining the $\Delta \omega$ range given the interaction
time $t_{int}$, $\Delta \omega =\omega_{\bf{k}}
+{\omega_{\bf{k}}}^{\prime}-\omega _P$, $\Delta {\bf k=k+k} ^{\prime
}-{\bf k}_P$.
For $t_{int}\rightarrow \infty$, $T(\Delta \omega)\rightarrow \pi \delta(\Delta \omega)$.
\bea
\label{PsiTilde}
 \widetilde{\psi }_{lp}\left( \Delta {\bf k}\right)
=\int_{V_I} d^3r\:\psi _{lp}\left( {\bf r}\right) \exp \left(
-i{\Delta }{\bf k}\cdot {\bf r} \right) \:,
\eea
 and $\psi _{lp}$ is the
field amplitude in ${\bf E}_P\left( \rho ,\phi ,z;t\right) = \psi
_{lp}\left( {\bf r}\right) e^{i\left( k_Pz-\omega _P t\right) }
\hat{{\bf e}}$.

$\widetilde{\psi }_{lp}\left( \Delta {\bf k}\right)$ is an effective transform of the pump field modulated by the phase term $\exp \left(-i{\Delta }{\bf k}\cdot {\bf r}\right)$ over the illuminated $(x,y,z)$  volume.  Due to the finite volume, there is a strong restriction in the $z$ variable, keeping the interaction within the crystal thickness $l_c$.  The mild simplification $l_c/z_R\ll 1$ gives \cite{Barbosa-WFunction}
\bea
\label{psiLargezR}
\!\!\!\!\widetilde{\psi }_{lp}\left( \Delta {\bf k},\xi \right)&=&A_{lp}\pi l_c (-1)^l  \left[\frac{k_P}{2 z_R} \right]^{(l/2)-1}\frac{(l+p)!}{l! p!}
\nonumber\\
&&\!\!\!\!\hspace{-2cm}\times \frac{ \sin\left[ l_c (\Delta k_z/2)\right]  }{  \left[ l_c (\Delta k_z/2)\right]  }\xi^{l/2} \mathcal{G}^p_l(\xi)\nonumber\\
&& \!\!\!\!\hspace{-2cm}\times e^{i \left( l \pi/2 +  l_c \Delta k_z/2-l \arctan(\Delta k_y/\Delta k_x)-(l_c+2 z_0)\Delta k_z/2 \right)}\:,
\eea
where $\xi$ is the two-point  transverse coordinate
\bea
\xi\!=\!\frac{z_R}{k_P} \left(\Delta k_x^2+\Delta k_y^2\right)\!\!=\!\!
\frac{z_R}{k_P}\left(\rho^2 +\rho'^2+2 \rho \rho'\cos (\phi-\phi')   \right),
\eea
and $\Delta k_j$
are cartesian components of $\Delta {\bf k=k+k} ^{\prime
}-{\bf k}_P$. $z_0$ gives the $z$ position for the minimum waist (focus) of the pump beam.
 $\rho=k \sin \theta,\rho'=k'\sin \theta'$.  $z_R$
is the Rayleigh range.

$\mathcal{G}^p_l(\xi)$ is the {\em polynomial}  of order $l$ defined by
\bea
\label{poly}
\mathcal{G}^p_l(\xi)\equiv\sum^{\infty}_{q=0} \frac{\left(-\xi/2\right)^q}{q!} \:
_2\mathcal{F}_1\left( -l,1+l+q;1+p;2\right)\:,
\eea
 and $_2\mathcal{F}_1$ is the hypergeometric function (The series (\ref{poly}) reduce to polynomials for specific $l$ values).

One should observe that the angle $\varphi_k$ appearing in $e^{- i l \arctan(\Delta k_y/\Delta k_x)}\equiv e^{-i l \varphi_k}$, in Eq.~(\ref{psiLargezR}), is a global angle connected with the signal {\em and} idler photons. This angle is a main spatial signature of the laser mode.
For a degenerate case and $k=k'$, $\theta=\theta'$ it is easy to see that  $\varphi_k\simeq \pi/2$.

One may write Eq.~(\ref{pm}) as
\bea
&&- i  \int_{t-t_{int}}^{t} H dt\nonumber \\&&\!\!\!=
\kappa\left( \Delta {\bf k},\xi \right) \:
{a}^{\dagger}\!\left( {\bf k} \right)
{b}^{\dagger}\!\left( {\bf k}^{\prime}\right)
 \!\!+\kappa^*\left( \Delta {\bf k},\xi \right) \:
{b}\!\left( {\bf k}^{\prime}\right) \left({a}\!\left( {\bf k} \right)
\right),
\eea
where
\bea
\kappa\left( \Delta {\bf k},\xi \right)&=&- i\sin \theta'\: k^{\prime 2}\:\sin \theta \:k^2\:\chi _{qrs}
\left( {\bf e}_{   {{\bf k},pm}            }\right)_r^{*}
\left( {\bf e}_{   {{\bf k}^{\prime},pm} }\right)_s^{*}\:\nonumber\\&&\times
l_{  \omega_{{\bf k}}     }^{(*)}\:
l_{\omega_{{\bf k}}^{\prime}}^{(*)}
T(\Delta \omega) \:
\widetilde{\psi }_{lp}\left( \Delta {\bf k}\right)(\Delta k)^6
\:.
\eea
$\kappa$ is an effective efficiency {\em function}. Frequently, only a numerical value giving its order of magnitude is used but, in fact, it contains all details of PDC. $\widetilde{\psi }_{lp}\left( \Delta {\bf k}\right)$ carries details of the coincidence probability structure that can be seen by high-order imaging. See Figs. 6,7 and 8 in \cite{Barbosa-WFunction}. These patterns could  be chosen as the desired high-order image to be tested. In the same way, other choices, such as different laser mode modes or symbols created by optical masks modifying the laser mode, will generate distinct $\widetilde{\psi }_{lp}\left( \Delta {\bf k}\right)$.

Keeping in mind that the structural detail of the high-order image is within the function $\kappa$ (that is to say, the information imprinted in the laser beam mode and implicit in a generic $\widetilde{\psi }_{lp}\left( \Delta {\bf k}\right)$), one may work on
\bea
\label{intH}
- i  \int_{t-t_{int}}^{t} H dt=
\left(\kappa \:
{a}^{\dagger}
{b}^{\dagger}
 +
 \kappa^*
{b} \:a
\right)
\:.
\eea
For this work, the exponential operator in Eq.~(\ref{Psi})  was expanded up to the 10$^{th}$ order (just to show photon numbers above the neural noise) and applied to the initial state $|\Psi \rangle_{\tiny signal}\otimes |0 \rangle_{\tiny idler}$. The structure implicit in the function $\kappa$ will ignored for the moment and $\kappa$  will be be taken just as a numeric value representing its order of magnitude close to the phase matching max. Whenever the structure itself becomes the object under investigation, one has to use $\kappa$ with all of its details.

Going back to Eq.~(\ref{Psi}), a handy mode to be used as the stimulating source to intensify the down-conversion process is a Gaussian coherent state that represents a laser beam, that is to say, $|\Psi \rangle_{\tiny signal}=| \alpha\rangle$. The coherent state is given by \cite{Glauber}
\bea
\label{alphaState}
| \alpha\rangle=e^{-|\alpha|^2/2}\sum_{n=0}^{\infty} \frac{\alpha^n}{\sqrt{n!}} |n \rangle \:.
\eea
$\alpha$ will be taken as real from now on and the stimulating laser beam is to be considered a single mode plane wave as a simplification (just to avoid a wave packet formalism at this moment).

The coefficients of the kets in the number of photons $n_s$ and $n_i$ in the wave function $|\Psi\rangle=\sum_{n_s} \sum_{n_i} A(n_s,n_i) |n_s\rangle |n_i \rangle$, obtained by expansion (\ref{Psi}), are the amplitude probabilities for occurrences of $n_s$ and $n_i$ photons. The magnitude square of these amplitudes are the probability $P(n_s,n_i)=|A(n_s,n_i)|^2$ of occurrence of $n_s$ and $n_i$ photons.



\begin{thebibliography}{}
\bibitem{correlation}
By ``correlation'' among variables it is implied that their variations are mutually dependent.
\end{thebibliography}
\end{document}